\documentclass[a4paper,conference]{IEEEtran}

\usepackage{graphicx}
\usepackage{algorithm}
\usepackage{makecell}
\usepackage{algorithmicx}
\usepackage{multirow}
\usepackage{multicol}
\graphicspath{{figure/}}
\usepackage{color}
\usepackage{amsmath}

\DeclareGraphicsExtensions{.pdf,.jpeg,.png}

\begin{document}

\title{Multi-modal Datasets for Super-resolution}


\author{\IEEEauthorblockN{
Haoran Li\IEEEauthorrefmark{1},
Weihong Quan\IEEEauthorrefmark{1},
Meijun Yan\IEEEauthorrefmark{1},
Jin zhang\IEEEauthorrefmark{1},
Xiaoli Gong\IEEEauthorrefmark{1}and
Jin Zhou\IEEEauthorrefmark{2}}
\IEEEauthorblockA{\IEEEauthorrefmark{1}
College of Computer Science, Nankai University, Tianjin, China,\\
College of Cyber Science, Nankai University, Tianjin, China,\\
Tianjin Key Laboratory of Network and Data Security Technology, Tianjin, China\\}
\IEEEauthorblockA{\IEEEauthorrefmark{2}Tianjin Chest Hospital, China}}


\maketitle


\begin{abstract}

Nowdays, most datasets used to train and evaluate super-resolution models are single-modal simulation datasets. However, due to the variety of image degradation types in the real world, models trained on single-modal simulation datasets do not always have good robustness and generalization ability in different degradation scenarios. Previous work tended to focus only on true-color images. In contrast, we first proposed real-world black-and-white old photo datasets for super-resolution (OID-RW), which is constructed using two methods of manually filling pixels and shooting with different cameras. The dataset contains 82 groups of images, including 22 groups of character type and 60 groups of landscape and architecture. At the same time, we also propose a multi-modal degradation dataset (MDD400) to solve the super-resolution reconstruction in real-life image degradation scenarios. We managed to simulate the process of generating degraded images by the following four methods: interpolation algorithm, CNN network, GAN network and capturing videos with different bit rates. Our experiments demonstrate that not only the models trained on our dataset have better generalization capability and robustness, but also the trained images can maintain better edge contours and texture features.



\end{abstract}

\IEEEpeerreviewmaketitle




\section{Introduction}

Single image super-resolution (SISR) \cite{glasner2009super} is defined as a process of reconstructing high-resolution (HR) image from a low-resolution (LR) image. Due to the ill-posed nature of SISR process, how to ensure that the fine textures \cite{ledig2017photo,zhang2019image} can still be generated without changing the image information is still a challenge \cite{nasrollahi2014super}. LR images used to train SISR models are often obtained by directly downsampling HR images. Some SISR methods choose to add Gaussian noise to LR images to improve the generalization ability of the model \cite{singh2014super}, and the specific process is as follows:

\begin{equation}\label{CR}
I^{LR} = (I^{HR} \otimes k)	\downarrow_s+ n
\end{equation}

where $(I^{HR} \otimes k)$ represents the convolutions between a blur kernel $K$ an the HR image, where $\downarrow_s$ represents a downsampling operation with different scales, and $n$ represents that Gaussian noise is added to the down-sampled LR.

\begin{figure}[t]
\centering
\includegraphics[height=5.0cm, width=8cm]{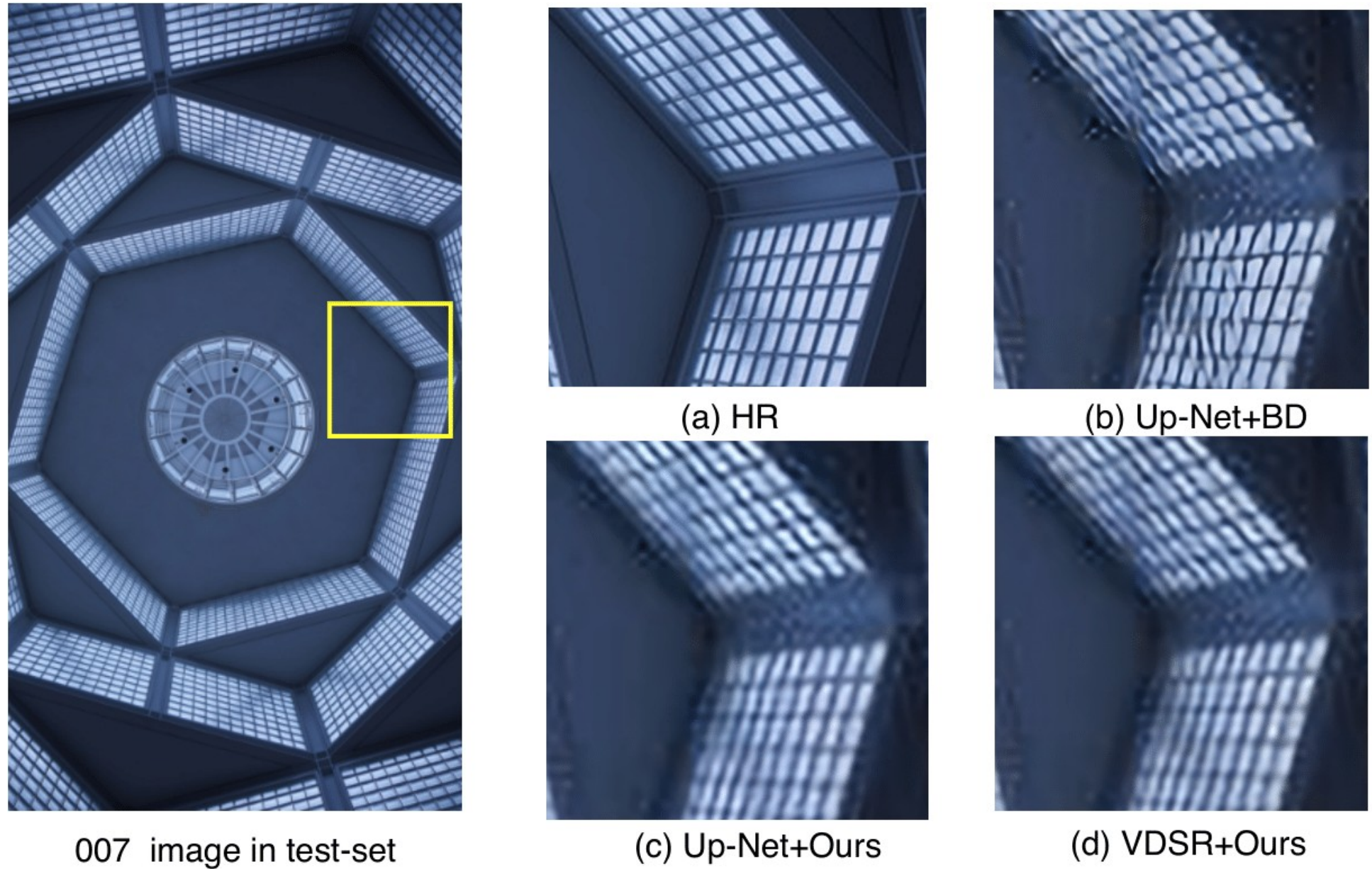}
\caption{The reconstruction result ($\times 4$) of NO.007 degraded image in our test set.
(a) the HR image, 
(b) image generated by Up-Net and trained on bicubic degradated image (BD),  
(c) image generated by Up-Net and trained on MDD400, 
(d) image generated by VDSR model and trained on MDD400.}
\label{007}
\end{figure}

Degradation of real-world images is often more complex, and complex factors such as motion, defocusing, compression, sensor noise need to be considered \cite{Yang2014Single,Romano2016RAISR,bulat2018learn}. Therefore, in the process of SISR, the model trained using the dataset obtained by the bicubic interpolation method is less robust in reconstructing the real-world image. The fundamental reason is that the bicubic downsampling algorithm used to build datasets are difficult to simulate the degradation process of real-world images. Fig. \ref{007} shows the SISR result of NO.007 degraded image in our MDD400 test set. We used VDSR \cite{kim2016accuratever} and UP-net (proposed by this paper) to train three models using simulated images with bicubic degradation and MDD400 which is proposed by this paper. The results illustrate that the model trained using the bicubic interpolation is difficult to solve the problem of reconstruction of degraded images, while the model trained on our dataset shows good performance.

In order to improve the robustness and generalization capability of SISR model for practical application. Two GAN networks are used to complete the super-resolution reconstruction of the image \cite{bulat2018learn}, one of which simulates the degradation process of real-world images as much as possible, thereby generating LR images closer to the real world. \cite{zhang2018learning} propose a simple yet effective and scalable deep CNN framework for SISR. The proposed model surpasses the widely-used bicubic degradation assumption and works for multiple and even spatially variant degradations, thus making a substantial step towards developing a practical CNN-based super-resolver.

 Therefore, compared to simulated LR and HR image pairs, a training dataset containing real-world data is needed. To the best of our knowledge, there is currently only one real-world-based\cite{cai2019toward} dataset constructed with photos of different resolutions by adjusting the camera's focus. However, the images in this dataset are all true-color photos, ignoring the  other type of images - black-and-white old photos, which are common in the museum or people's home. These photos tended to be low resolution due to the limitations of shooting and film development technology at the time. Repairing and enlarging them not only helps historians to better recover history, but also allows ordinary people to remember the past, which has a high scientific and social value. Therefore, we propose a dataset of old photos based on the real world. We took photos with a film camera and generated an electronic version of the old photos as LR images with a scanning device, while trying to take the photos of the same content in the same location as the corresponding HR image with a digital single-lens reflex (DSLR) camera Canon 650D. And then we designed the corresponding image alignment algorithm to obtain a precisely aligned image pair.


At the same time, for the images of degraded scenes in the real world, the ground truth of them is often difficult to obtain through photography, and can only be obtained through simulation. Therefore we propose the MDD400 that simulates real-world degradation. We first constructed a new dataset of 300 images as ground truth, and then used the bicubic interpolation algorithm to obtain the corresponding LR images to construct 100 pairs of interpolation degraded images. Next, we designed a GAN-based image enhancement method for the super-resolution reconstruction model. Up-Net upsampling was used to generate 100 pairs of CNN-class degraded images. Because the images generated by CNN are too smooth and the texture is not clear, we constructed texture-Net to generate corresponding texture details for the image to form 100 pairs of GAN-class degraded images. Finally, we captured 100 sets of videos with different bit rates to construct image pairs with different resolutions to form video degradation images.

The contribution of our work are threefold:
\begin{itemize}
\item We provide a real-world-based dataset OID-RW that includes character and architecture. The model trained on this dataset shows better results in practical applications.
\item We provide a dataset MDD400 that simulates dal degradation of real-world images, which include interpolation class, CNN class, GAN class and video class. When reconstructing real-world images, the model trained on this dataset has strong generalization capability and high robustness.
\item We propose a GAN-based data augmentation algorithm suitable for SISR, that generates LR images with different degrees of degradation by selecting different loss functions.

\end{itemize}

In section \ref{sec:relatedworks}, datasets and data enhancement algorithms applied in the super-resolution field are introduced.
Section \ref{sec:real-world dataset} and section \ref{sec:Multimodal dataset} describe the construction process of  real-world-based SISR dataset and the multi-modal degradation dataset.
The last section \ref{sec:experiment} shows the experimental details and result analysis.
\section{Related Work}
\label{sec:relatedworks}

\subsection{Singal Image Super-resolution Datasets}
 The datasets commonly applied in SISR training so far primarily are DIV2K \cite{timofte2017ntire}, Urban100 \cite{huang2015single}, BSD300 \cite{martin2001database}. Set5 \cite{bevilacqua2012low} and Set14 \cite{zeyde2010single} are often used for model evaluation. The model trained by the dataset which is generated through the bicubic interpolation algorithm lacks the generalization capability. In recent years, a lot of work has been done to make SISR models perform well in real-world scenarios. Qu et al. \cite{Qu2016Capturing} presents a novel prototype camera system to address the aforementioned difﬁculties of acquiring ground truth SR data. Two identical camera sensors equipped with a wide-angle lens and a telephoto lens respectively to collect a face image dataset. Köhler et al. \cite{KToward} introduce the first comprehensive laboratory SR database of all-real acquisitions, which consists of more than 80k images of 14 scenes combining different facets. Cai et al. \cite{cai2019toward} captured images from various scenes at multiple focal lengths, providing a general and easy-to-use benchmark for real-world single image super-resolution. chen et al. \cite{ChenCamera} captured City100 dataset at on one scaling factor. The above datasets often focus on true color images, but SISR work on black and white photos of film cameras is still of great practical significance, so this paper builds a dataset based on old photos.
 
 
 \subsection{Image Degradation}
 
In recent works, in order to improve the generalization ability of the model, blur kernels and noise have been added to LR images. These two factors have been recognized as key factors for the success of SISR and several methods have been proposed to consider them. zhang et al. \cite{zhang2018learning} propose a dimensional stretching strategy that takes blur and noise as inputs. This method can cope with multiple and spatially changing degradation models, which obviously improves the practicality. Adrian et al. \cite{bulat2018learn} use a mismatched HR and LR image training set to train a GAN network that converts HR images into LR images, and then use this network to simulate the generation of degraded images in the real world. We are the first to propose a multi-modal image degradation dataset. Based on previous work, we used CNN and GAN networks to generate degraded images. In order to expand the diversity of the dataset, we captured videos at different bit rates.



\section{Old Real-world Dataset}
\label{sec:real-world dataset}

This paper mainly constructs two types of images, including character and architecture. For character, HR images are obtained by manually filling pixels. For architecture, HR and LR images are obtained by shooting with two kinds of cameras. The dataset contains 82 groups of images, of which 22 are character and 60 are architecture. Although the number of photos seems to be small, the size is large enough to fit the quantity level of the training model.

\begin{figure}
\centering
\includegraphics[height=10.0cm,width=8cm]{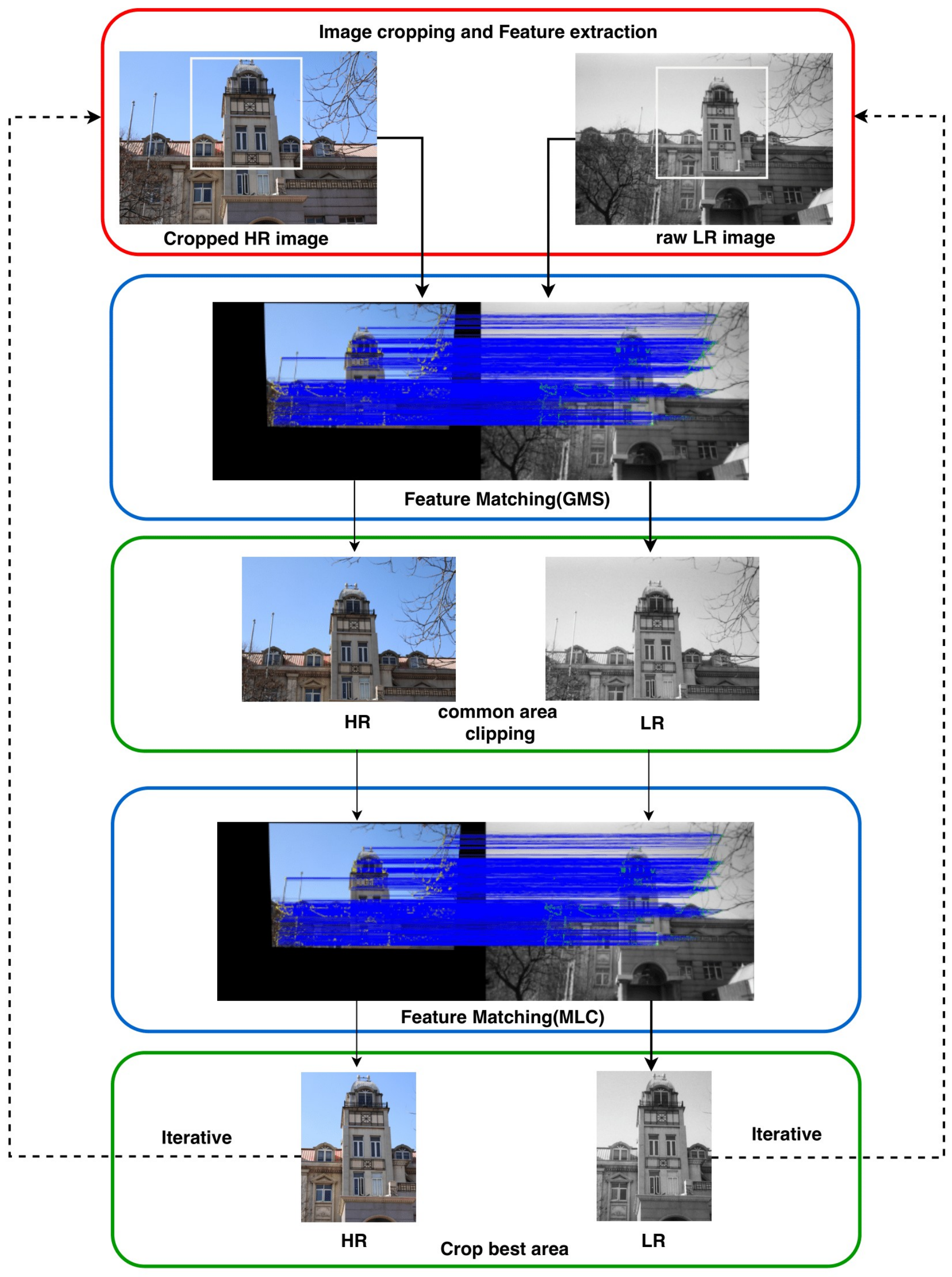}
\caption{Image alignment process.}
\label{real-world}
\end{figure}

\subsection{Image Collection}

 We took the same picture scene at the same position and angle as far as possible, using film camera and CCD camera sharing the same focal length and other parameters, to construct the dataset. The LR images were constructed by digital scanning format of photographic film which were developed by adopting professional reagents and instruments in the darkroom. Correspondingly, the HR images were ideally built by using CCD cameras. However, due to the different image formations of the two cameras and other uncontrollable factors such as illumination and wind, it was impossible for the two cameras to correspond completely and there was doomed to be some deviations. The paper applies an approach aligning the two kinds of images by a series of image processing methods.

\subsection{Image Registration}
\label{sec:align}

As mentioned earlier, the images of the same scene taken by two cameras cannot be completely aligned at the pixel level. Thus, we design an image alignment algorithm to process the captured images, so as to obtain the precisely aligned image pairs. We perform the steps of image cropping, feature extraction, feature matching, image affine transformation and clipping of the common area based on the obtained image pairs, as is shown in Fig. \ref{real-world}.

\textbf{Image cropping and Feature extraction.} We first cropped the invalid part of the image, such as pedestrians, shaking branches, waving red flags, moving cars and so on. As to the feature extraction, we make a selection among the SUFT \cite{bay2006surf}, ORB \cite{rublee2011orb} and SIFT \cite{lowe2004distinctive} algorithms. We chose SIFT algorithm because robustness rather than time is mainly focused during the construction of dataset.

\textbf{Feature Matching.} First of all, we manage to get as many matching pairs as possible. After optimizing the selected feature matching pairs by GMS and MLC, we retain the correct matches and removed the wrong matches as possible as we can. The main idea of GMS \cite{bian2017gms} optimization is to judge whether there are multiple support matches around the correct match. 

The main idea of matching location constraints (MLC) is that the same feature points in the image should be roughly the same, which means there is no need to consider the rotation consistency and other issues. The size of image A and image B is $M \times N$. For a correct match $r_i$, the position of the feature point in image A is $P_A(x_r^A, y_r^A )$, and the position of feature point in image B is $P_B(x_r^B, y_r^B )$, and the two feature points should meet the constraints:

\begin{equation}\label{CR}
\newcommand * \abs[1]{\lvert#1\rvert}
\abs{x_r^A - x_r^B } \le t_x, \quad t_x > 0 
\end{equation}

\begin{equation}\label{CR}
\newcommand * \abs[1]{\lvert#1\rvert}
\abs{y_r^A - y_r^B } \le t_y, \quad t_y > 0 
\end{equation}

Where $t_x$ is the threshold in the x-axis direction and $t_y$ is the threshold in the y-axis direction. It satisfied
${t_x}/{t_y} = {M}/{N} $ and $t_x = \alpha M $, $\alpha \in (0, 1)$.

The smaller the $ \alpha $, the smaller the threshold $ t $. On the contrary, the larger the $\alpha$, the higher the probability of removing the mismatch. For the two images have already been conducted with an alignment process using matching location constraints, and our following images alignment have limited acceptance of error matching, the value of $\alpha$ in this paper will be set smaller. We set $\alpha = 0.1$ for our dataset.


\textbf{Affine transformation and common area clipping.} According to the feature point matching pairs obtained in the previous step, we complete the viewpoint matching of raw-LR and raw-HR images by calculating the affine variation matrix. We introduce the RANSAC algorithm \cite{fischler1981random} to reduce the effect of noise (error matching point pairs) on the image alignment in the calculating process. After the image alignment operation, we got the raw-HR images with black edges. We perform the global binary process to those images to extract the edge contour, and then design an algorithm to detect and capture the largest inscribed quadrilateral as the clipping region of the final HR images. At this time, the raw-LR image is also clipped according to the four corresponding angular coordinates of the inscribed quadrilateral to get the final LR image.

\section{Multi-modal Dataset}
\label{sec:Multimodal dataset}

\begin{figure*}[t]
\centering
\includegraphics[height=7.0cm,width=17cm]{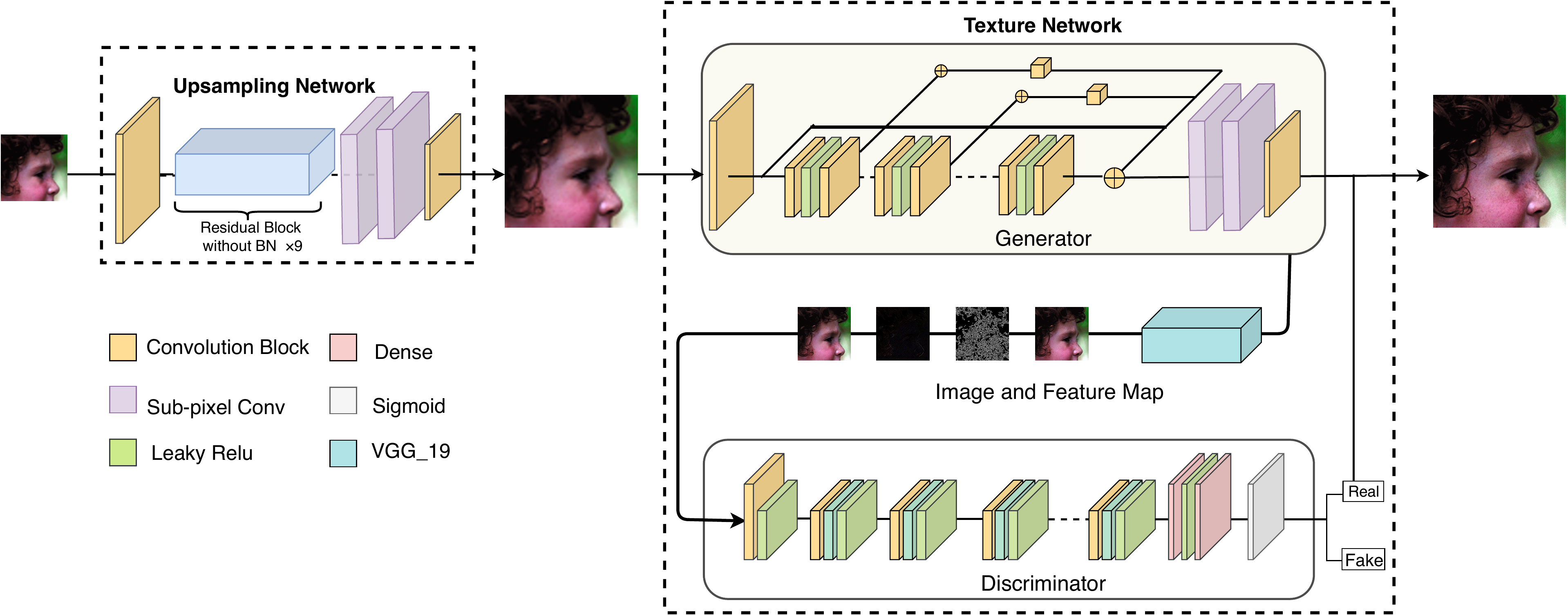}
\caption{Overall proposed architecture and training pipeline.}
\label{pipeline}
\end{figure*}



For better super-resolution results in real scenarios, we propose a new dataset NRD300 and a new data augmentation pipeline to synthesize more realistic training data. We build a multi-modal image degradation dataset MDD400 on the basis of NRD300 by using CNN and GAN network simulation to generate degraded images. In order to increase the diversity and generalization capability of the dataset, we captured videos of different resolutions as part of MDD400.



\subsection{New Resolution dataset}
\label{sec:NRD300}

We first build a new dataset called New Resolution Dataset 300 (NRD300) for SR domain. By analyzing the dataset used to train SR model, we find that although the number of the images is quite large, they are not rich in species. Therefore, in order to build a comprehensive dataset in SR field, we select the images of ancient Chinese architecture and cultural relics to increase the diversity of images for the current dataset. After that, we use our new dataset NRD300 to build a multi-modal images degradation dataset called MDD400. 

\subsection{Degraded Image by CNN}
\label{sec:CNN}

\begin{figure}
\centering
\includegraphics[height=4.0cm,width=8cm]{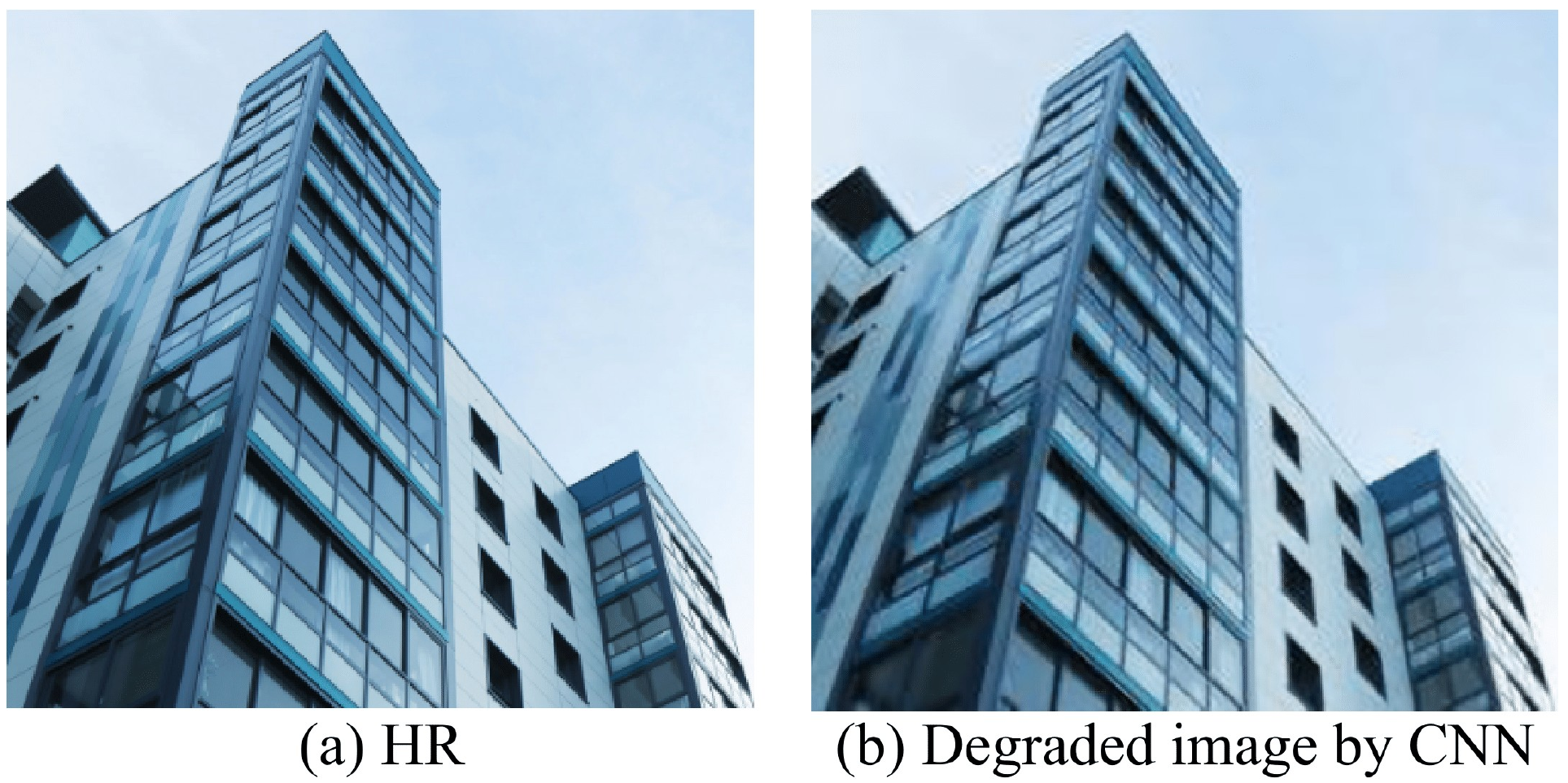}
\caption{Degraded images by CNN networks.}
\label{degradtion_CNN}
\end{figure}

The Up-Net model based on CNN is applied to upsample the LR images. Fig. \ref{pipeline} shows overall proposed architecture and training pipeline. We use a lightweight network to perform super-resolution reconstruction of the LR. First, in order to better retain the texture information of the images, we apply 9 $\times$ 9 convolution to obtain the low-level features of the images, utilize the Residual Block to obtain the high-level features of the images, and use long-range skip connection to enhance the feature propagation in deep networks \cite{tong2017image}. In order to improve the training speed, we remove the BN layer from the Residual Block \cite{lim2017enhanced}, and then set the Residual Block to 9 blocks. Finally, the up-sampling of the image was completed through two layers of sub-pixel convolution \cite{shi2016real}. In the upsampling process, we use Mean square error (MSE) as loss function. The result is show in Fig. \ref{degradtion_CNN}.

\begin{equation}\label{CR}
\mathcal{L}_{MSE} = \parallel{}I^{HR}-I^{SR}\parallel{_2}
\end{equation}
where $I^{SR} $ is the output image of the Up-Net model, $I^{HR} $ is the HR image corresponding to $ I^{SR}$.

\begin{figure*}[t]
\centering
\includegraphics[width=16.5cm]{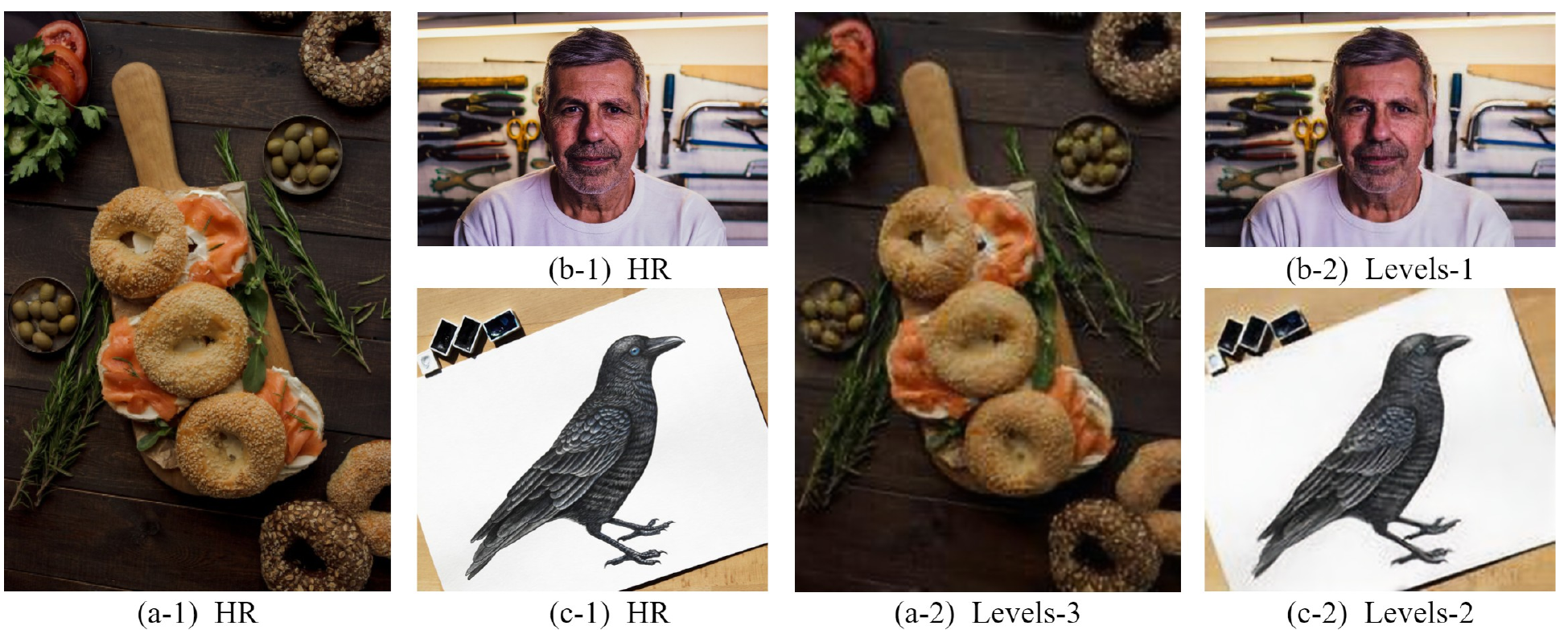}
\caption{Images of different degradation levels are generated by GAN.}
\label{deg-gan}
\end{figure*}

\subsection{Degraded Image by GAN}
\label{sec:Gan}


Inspired by \cite{bulat2018learn}, GAN can be used to effectively simulate the image degradation, thus we designed a GAN network called Texture-net to generate different levels of low-resolution degraded images. Texture-net adopts a consistent network structure of the Up-net, and the input is $I^{SR}$ which is obtained using Bicubic downsampling. Inspired by SRFeat \cite{park2018srfeat}, we apply $1\times1$ to each long-rang skip connection to adjust and balance the output.

In order to generate different levels of degraded images, we use multiple loss functions to train the network. Due to the rich texture information in the images generated by GAN, we use two discriminators for the authenticity and fineness of the generated textures. One is the image discriminator, which is used to discriminate the authenticity of the generated images. The other is a perceptual discriminator, which extracts the advanced features of the image through VGG-19, and judges the authenticity of the generated image features \cite{radford2015unsupervised}. We use the following four loss functions to train the Texture-net.


\textbf{Image Adversarial loss.} The GAN loss of image is defined as follows:
\begin{equation}\label{CR}
\mathcal{L}_{adv} = -\log\, [\mathcal{D}_i\, (I^{tSR})\, ]
\end{equation}
\begin{equation}\label{CR}
{L}_{d} = -\log\, [\mathcal{D}_i\, (I^{HR})\, ] -\log [1-\mathcal{D}_i\, (I^{tSR})]
\end{equation}
where $ I^{tSR}$ is the output image of the Texture-Net model, $I^{HR} $ is the high resolution image corresponding to $ I^{tSR}$, $\mathcal{L}_{adv}$ is the loss function of generator \cite{goodfellow2014generative}, ${L_d}$ is the loss function of discriminator, and $\mathcal{D}_i$ is the image discriminator. In our model we minimized $ \mathcal{L}_{adv}$.

\textbf{Feature Adversarial loss.} VGG-19 is used to extract the feature map of $I^{HR}$ and $I^{tHR}$, then the feature map is fed into feature discriminator to get real texture.
\begin{equation}\label{CR}
\mathcal{L}_{fadv} = -\log\, [\mathcal{D}_f\, (\phi(I^{tSR}))\, ]
\end{equation}

\begin{equation}\label{CR}
{L}_{d} = -\log\, [\mathcal{D}_f\, (\phi(I^{HR}))\, ] -\log [1-\mathcal{D}_f\, (\phi(I^{tSR}))]
\end{equation}
where $ \mathcal{L}_{adv}$ is the loss function of generator, ${L_d}$ is the loss function of discriminator, and $\mathcal{D}_f$ is the feature discriminator. 

\textbf{Perceptual loss.} Perceptual loss \cite{isola2017image} is often used for the GAN models to generate images with better visual quality, and we adopted the relu5\_4 layer of VGG-19.
\begin{equation}\label{CR}
\mathcal{L}_{per} = \frac{1}{W \times H}\sum_{i=1}^{C}\parallel{} \phi (I^{HR})-\phi({I^{tSR}})\parallel{_2}
\end{equation}
where $W$ is the width of feature map, $H$ is the height of feature map, $C$ is channel of the feature map, and $\phi$ is the feature map of VGG-19 on relu5\_4.

\textbf{Style loss.} Style loss is defined by the Gram matrix. To further make the texture generated by $I^{tHR}$ more realistic and close to $I^{HR}$, we applied the style loss function to neural style learning.
\begin{equation}\label{CR}
\mathcal{L}_{sty} = \frac{1}{W \times H}\sum_{i=1}^{C}\parallel{}\mathcal{G}[\phi (I^{HR})]-\mathcal{G}[\phi (I^{tHR})]\parallel{_2}
\end{equation}
where $\mathcal{G}$ is Gram matrix, $W$ is the width of feature map, $H$ is the height of feature map, $C$ is channel of the feature map, and $\phi$ is the feature map of VGG-19 on relu5\_4.

In order to increase the diversity of image in the dataset, we generated 3 levels of degraded images according to the number of the selected loss functions which is shown in Table \ref{tbl:table2}.
The result as show in Fig. \ref{deg-gan}.

\begin{figure*}[t]
\centering
\includegraphics[width=18cm]{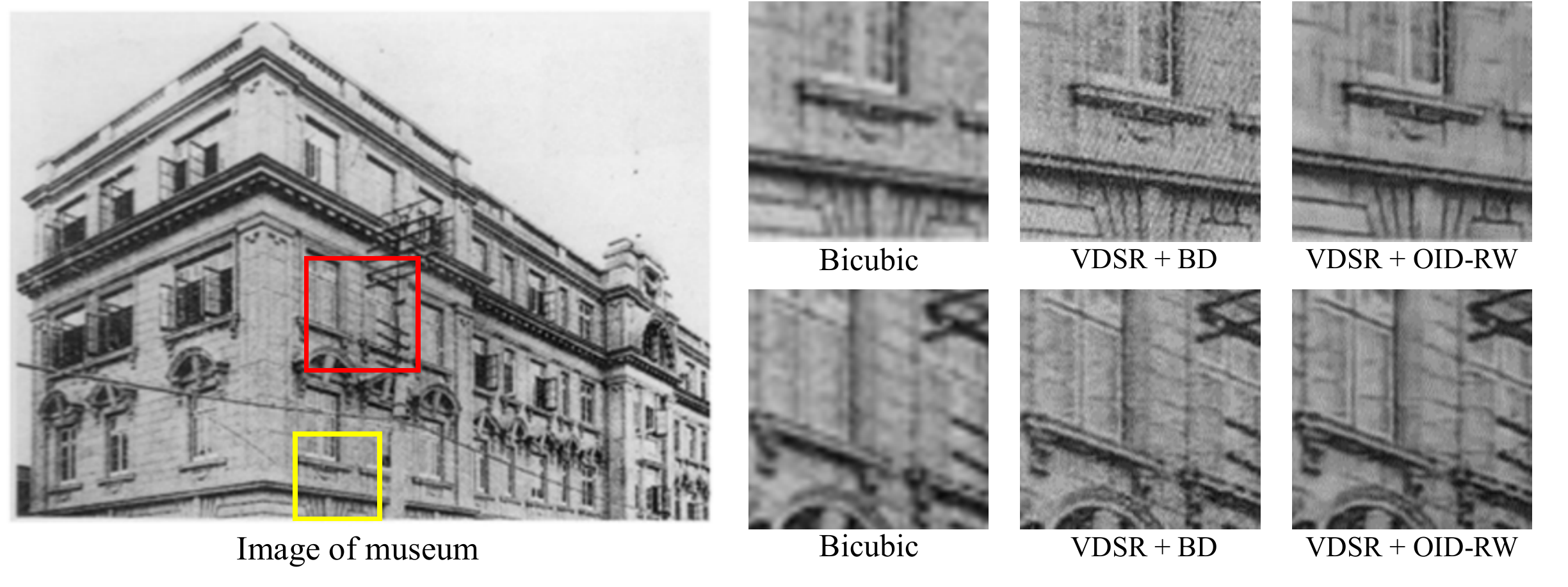}
\caption{SR results ($\times$4) on museum image by VDSR model trained on BD and OID-RW.}
\label{OID-RW}
\end{figure*}

\begin{table}[h]
\caption{Images of different degradation levels are generated by GAN.}
\label{tbl:table2}
\begin{center}
\begin{tabular}{|c||c|}
\hline
Levels & Loss Fuctions \\ 
\hline
1 & $\mathcal{L}_{adv}+\mathcal{L}_{fadv}+\mathcal{L}_{per}+\mathcal{L}_{sty} $\\
\hline
2 & $\mathcal{L}_{adv}+\mathcal{L}_{fadv}+\mathcal{L}_{per}$ \\
\hline
3 & $\mathcal{L}_{adv}+\mathcal{L}_{fadv}$ \\
\hline
\end{tabular}
\end{center}
\end{table}

\subsection{Degraded Images from videos}
\label{sec:video}

We select a 1080P documentary and its 360P version on a video website, and then play them in a same size screen to capture the video at the same time. To obtain more image pairs, we also select other types of videos. In the process of capturing, we try to capture the still parts to obtain the aligned image pairs. However, for the reason that there are many frames of video in one second, the captured images may not be fully aligned. Therefore, we used the alignment algorithm of Section \ref{sec:align} to align them.

\begin{figure}[h]
\centering
\includegraphics[height=4.0cm,width=8cm]{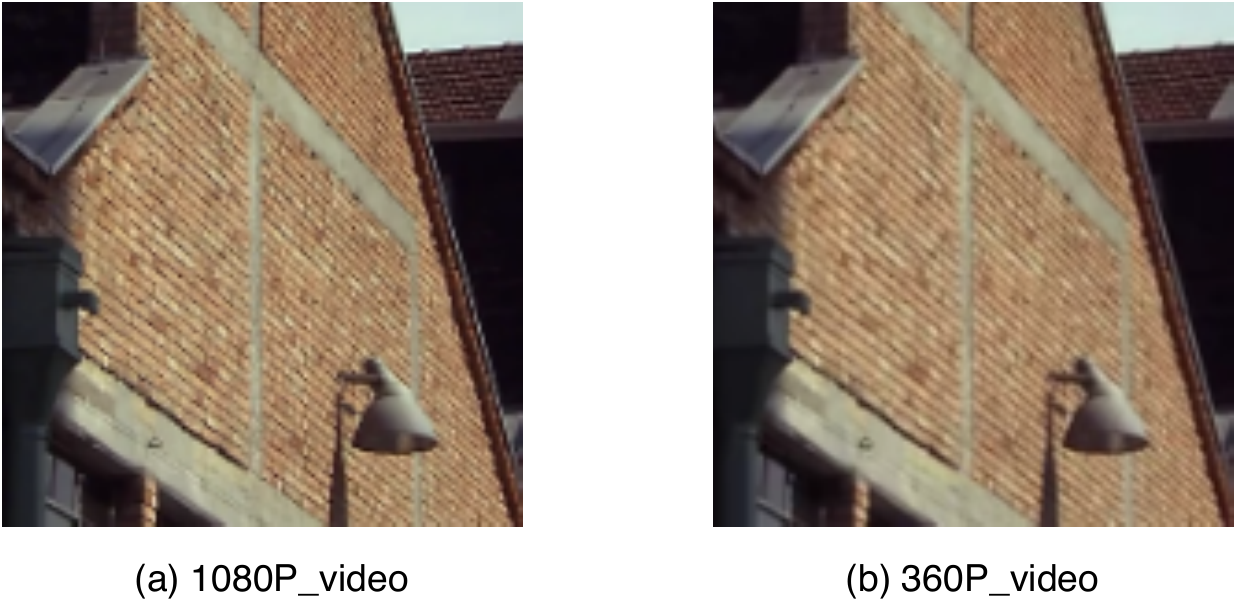}
\caption{Degraded images from different videos}
\label{video}
\end{figure}

\section{Experiments and Results}
\label{sec:experiment}

\subsection{Training Details}

\textbf{Training process.} Here are the details of training Up-Net and Texture-Net networks. We use the training set and test set of DIV2K as the training set of our model (a total of 900 images). We randomly select 500 training images and generate LR images by bicubic interpolation as the input of Up-Net. Data augmentation was performed by randomly rotating $90^{\circ}$, $180^{\circ}$, $270^{\circ}$ degrees and horizontally flipping the input. After training the UP-Net network, the remaining 400 images after downsampling are fed into the Up-Net to get $I^{SR}$. $I^{SR}$ and the corresponding HR form the training set of Texture-Net. During the training process, Up-Net and Texture-Net are both trained for 20 epochs. The experimental environment for OID-RW is Intel Core-i7, Nvidia GeForce GTX 1080ti, Nvidia titan xp and 32G memory. Experiments on MDD are conducted on NVIDIA Tesla P100.

\textbf{Parameter settings.} In order to balance the different loss functions, the feature map is scaled with scaling factor, and the weights of $\mathcal{L}_{adv}, \mathcal{L}_{fadv}, \mathcal{L}_{per}$ and  $\mathcal{L}_{sty}$ are set to $10^{-3}$, 1, $10^{-3}$, 1 respectively. The momentum parameter $\beta$ of Adam optimizer is 0.9.

\begin{figure*}[t]
\centering
\includegraphics[width=14cm]{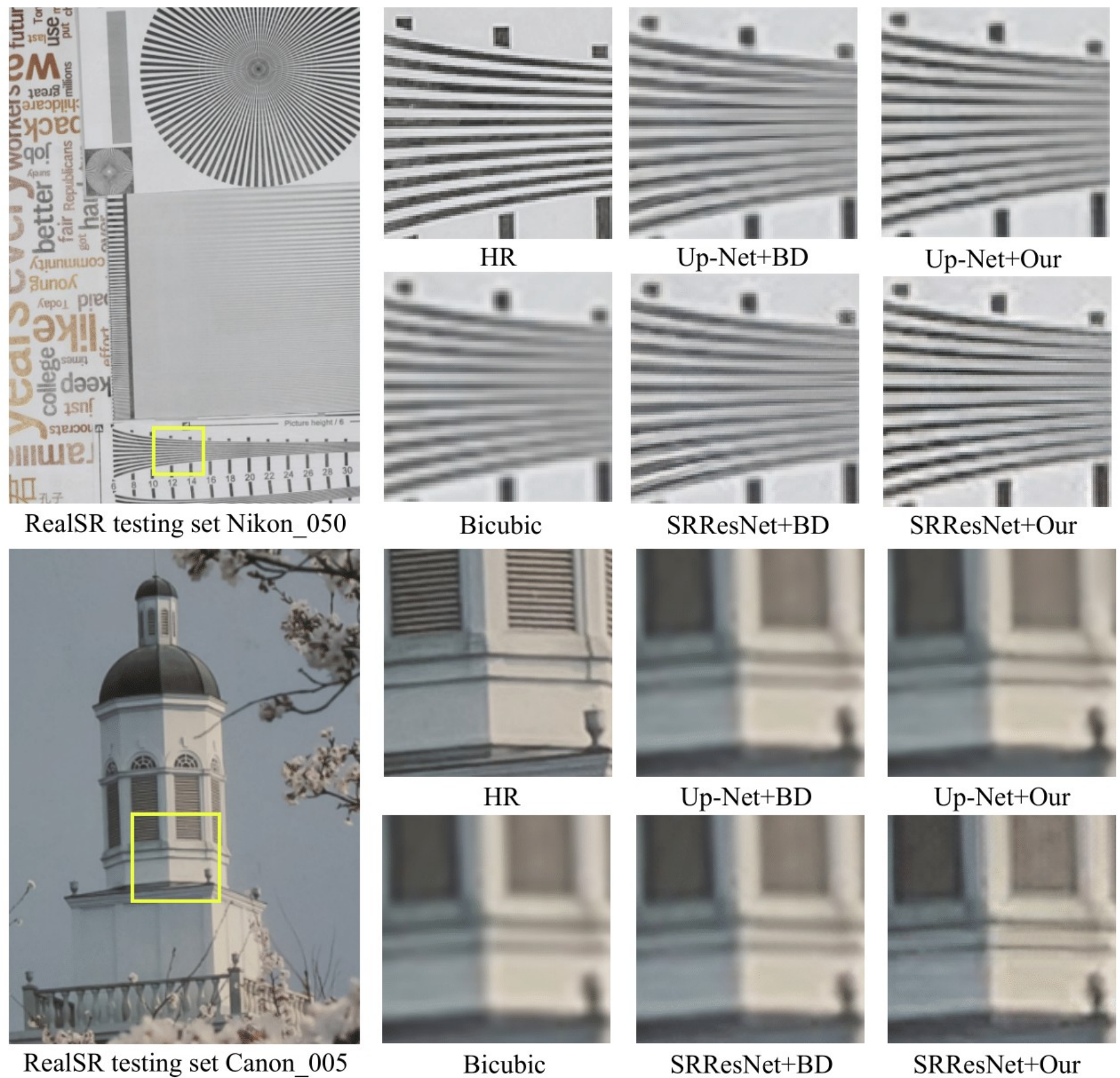}
\caption{SR results ($\times$4) on RealSR test set\cite{cai2019toward} by different methods trained on BD and MDD400.}
\label{Nikon}
\end{figure*}

\subsection{OID-RW Dataset vs. Simulated Dataset}

We choose the state-of-art SISR model called VDSR \cite{kim2016accuratever} and treat its performance on different datasets as the evaluation index to demonstrate the advantages of OID-RW dataset. In order to speed up the overall training, we reduce the number of VDSR intermediate layers (which is set to 15). Then we utilize the original dataset used by VDSR to obtain the training set BD by bicubic interpolation downsampling. The OID-RW dataset are divided into a training set and a test set at a ratio of 8: 2. The input image are shuffled with factor $1/2$, $1/3$, $1/4$ for the three scaling factors $\times2$, $\times3$ and $\times4$, respectively. 

Finally, the VDSR model is separately trained using OID-RW and BD training set, and tested on the OID-RW test set. PSNR and SSIM (on Y channel in the YCbCr space) are used to evaluate the performance of the model as Table \ref {coarse_grained_model_eval_result} shows. Results show that VDSR trained on our OID-RW dataset obtain significantly better performance than that trained on BD dataset for all three scaling factors. Specifically, for scaling factor 3 and 4, VDSR trained on our OID-RW dataset has about 0.53dB improvement on average for different magnification.



\begin{table}[h]
\centering
\caption{Evaluation results of coarse-grained SR models.}
\label{coarse_grained_model_eval_result}
\resizebox{0.48\textwidth}{!}{
\begin{tabular}{|c||c||c||c||c|}
\hline
\multirow{2}{*}{Magnification} & \multicolumn{2}{c||}{Trained with original dataset} & \multicolumn{2}{c|}{Trained with OID dataset} \\  \cline{2-5}
 & PSNR & SSIM & PSNR & SSIM \\
 \hline
2 & 25.58 & 0.76 & 26.01 & 0.82 \\
\hline
3 & 25.17 & 0.71 & 25.75 & 0.82 \\
\hline
4 & 24.78 & 0.62 & 25.36 & 0.81 \\
\hline
\end{tabular}}
\end{table}

\subsection{Generalization Capability of OID-RW Dataset}

\begin{figure}[h]
\centering
\includegraphics[width=9cm]{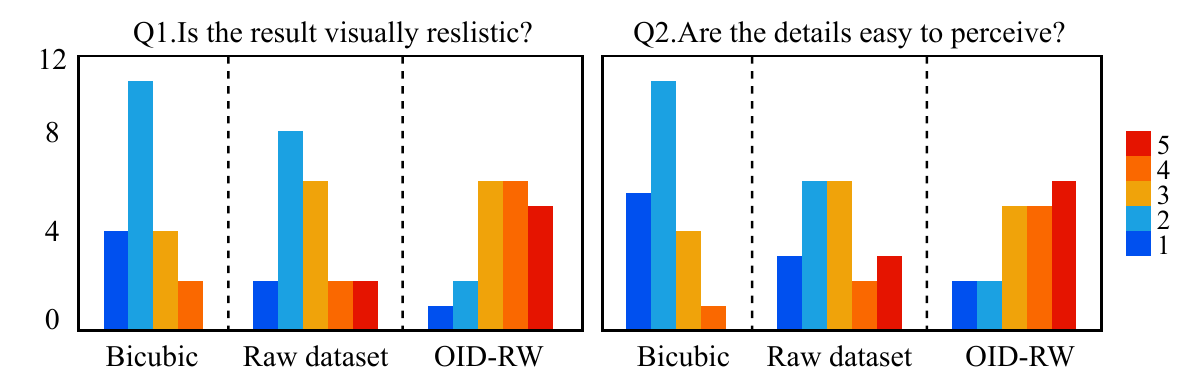}
\caption{Rating distributions for various methods on the two questions. The ordinate axis shows the rating frequency received by the methods from the volunteers}
\label{Subjective evaluation}
\end{figure}

In order to further prove the generalization capability and robustness of the OID-RW Dataset, this paper performs $\times2$, $\times3$ and $\times4$ super-resolution reconstruction on 10 museum photos respectively, and the super-resolved results are visualized as Fig. \ref{OID-RW}. Since there are no corresponding ground truth images of real museum photos, and objective evaluation indicators cannot be used, we adopted the method of manual scoring, inviting 20 volunteers for subjective evaluations. For each super-resolved result, the volunteers are asked to give a rating for two questions. One is \textit{Is the result visually realistic?}, and the other is \textit{Are the details easy to perceive?}. The Likert scale of the reconstructed image ($\times 4$) is shown in Fig. \ref{Subjective evaluation}, where the best score is 5, and the worst is 1.

It can be seen from the Fig. \ref{OID-RW} that in all these cases the images generated by VDSR trained on our OID-RW Dataset have better texture details and clear edge. The distribution shows that our results are more preferred, where our images receive more red and far less blue ratings compared to the others. As can be seen from Fig. \ref{Subjective evaluation}, the images trained on the OID-RW dataset have a better visual effect, and the texture information of the images is also more refined.

\subsection{SISR Models Trained on MDD400 Dataset}

In order to prove that the model trained on our MDD400 dataset can get good performance in degraded scenarios, the following experiments are designed. DIV2K is often selected as the training set for state-of-the-art SISR models. Here we use the bicubic interpolation algorithm on DIV2K to obtain a low-resolution dataset (BD), we train three networks including Up-Net, VDSR and SRResNet on the BD and MDD400 training sets.

We tested all the SISR models on our MDD400 test set. We evaluate the images in two scaling factors, $\times 2$ and $\times 4$, by PSNR and SSIM on the Y channel. The evaluation results are shown in Table \ref{tb2:PSNR}. Obviously, models trained on MDD400 dataset perform better than baselines and the models trained on BD dataset on all the factors. The models trained on our MDD400 dataset have about 0.8dB improvement on average for all the three SISR models, specifically for scaling factor $\times 2$.
 
\begin{table}
\caption{Comparison of Bicubic, Up-Net, VDSR \cite{kim2016accuratever}, SRResNet \cite{ledig2017photo} trained on BD and our MDD400 dataset.}
\label{tb2:PSNR}
\begin{center}
\begin{tabular}{|p{0.6cm}<{\centering}|p{0.5cm}<{\centering}||p{0.8cm}<{\centering}||p{0.4cm}p{0.5cm}||p{0.4cm}p{0.5cm}||p{0.4cm}p{0.5cm}|}
\hline
\multirow{2}{*}{Metric} &
\multirow{2}{*}{Scale} &
\multirow{2}{*}{Bicubic} &
\multicolumn{2}{c||}{Up-Net} &
\multicolumn{2}{c||}{VDSR} &
\multicolumn{2}{c|}{SRResNet} \\
\cline{4-9}
& & & BD & Ours & BD & Ours & BD & Ours\\
\hline 

\multirow{2}{*}{PSNR} & $\times 2$ &31.57 & 32.10 & \textbf{33.25}&32.35 & \textbf{33.54}& 32.67&\textbf{33.75} \\

\multicolumn{1}{|c|}{} & $\times 4$ & 27.45 &28.78 &\textbf{29.26} & 29.13&\textbf{29.62} &29.42 & \textbf{29.87}\\

\hline 
\multirow{2}{*}{SSIM} & $\times 2$ &0.877 & 0.903& \textbf{0.915}& 0.907&\textbf{0.918} &0.911 &\textbf{0.920} \\

\multicolumn{1}{|c|}{} & $\times 4$ & 0.796& 0.803&\textbf{0.812} &0.805 &\textbf{0.819} &0.803 & \textbf{0.820}\\
\hline
\end{tabular}
\end{center}
\end{table}
 
\subsection{Generalization Capability of MDD400 Dataset}
 
In order to prove that the model trained on MDD400 dataset has good robustness, we selected the test image NiKon\_050 and Canon\_005 in the RealSR test set for further experiments. We use BD and our MDD400 dataset to train Up-Net and SRResNet\cite{ledig2017photo} networks. From the Fig. \ref{Nikon} it can be seen that the image generated by the BD-trained model has a certain degree of distortion, but that generated by MDD400-trained model can generate clear edges profile.

\section{Conclusion}

The traditional SISR datasets are often obtained by bicubic interpolation of high-resolution images to generate low-resolution images, and sometimes noise are added after downsampling. Models trained on these datasets are not robust in practical applications. In this paper, we propose two datasets. One is based on the real-world old photo dataset OID-RW, and the other is the MDD400 dataset that simulates real-world multi-modal degradation. Our extensive experiments demonstrates that not only the real-world SISR results of models trained on our dataset is much better than those trained on existing simulated datasets, but also models trained on our dataset show better generalization capability.







\end{document}